\documentclass[11pt,preprint,nofootinbib,superscriptaddress]{aastex}

\long\def\dump#1{}

\def\beq{\begin{equation}}
\def\eeq{\end{equation}}
\usepackage{epsfig}

\def\iso#1#2{\mbox{${}^{#2}{\rm #1}$}}
\def\k4#1{\iso{K}{4#1}}
\def\u23#1{\iso{U}{23#1}}
\def\th23#1{\iso{Th}{23#1}}

\begin{document}

\title {Probing the Earth's interior with the LENA detector}

\author{Kathrin A.~Hochmuth}
\affil{Max-Planck-Institut f\"ur Physik
(Werner-Heisenberg-Institut), F\"ohringer Ring 6,\\ 80805 M\"unchen,
Germany}
\altaffiltext{}{email: hochmuth@ph.tum.de}

\author{Franz~v.~Feilitzsch, Teresa~Marrod\'an~Undagoitia, Lothar~Oberauer, \\Walter~Potzel, Michael~Wurm}
\affil{Technische Universit\"at M\"unchen,
Physik Department E15, James-Franck-Strasse, \\85748 Garching, Germany}

\author{Brian~D.~Fields}
\affil{Center for Theoretical Astrophysics,
Department of Astronomy, University of Illinois,
\\Urbana, IL~61801, USA}

\begin{abstract}
A future large-volume liquid scintillator detector such as the proposed 50 kton LENA (Low Energy Neutrino Astronomy) detector would provide a
high-statistics measurement of terrestrial antineutrinos originating
from $\beta$-decays of the uranium and thorium chains. Additionally, the neutron is scattered in the forward direction in the detection reaction
$\bar\nu_e+p\to n+e^+$. Henceforth, we investigate to what extent LENA can distinguish
between certain geophysical models on the basis of the angular
dependence of the geoneutrino flux.  Our analysis is based on a
Monte-Carlo simulation with different levels of light yield,
considering an unloaded PXE scintillator. We
find that LENA is able to detect deviations from isotropy of the
 geoneutrino flux with high significance. However, if only the directional information is used, the time required to distinguish between different geophysical models is of the order of severals decades. Nonetheless, a
high-statistics measurement of the total geoneutrino flux and its
spectrum still provides an extremely useful glance at the Earth's
interior.
\end{abstract}

\maketitle

\section{Introduction}

A future large volume liquid scintillator such as the proposed LENA detector (Oberauer {\it et al.~}2005) can obtain a high precision measurement of  the geoneutrino flux, could deliver new information about the interior of
the Earth, in particular its radiochemical composition, and thus give new insights on Earth and planetary formation.

Besides the geoneutrino measurement LENA will be designed for high-statistics solar neutrino spectroscopy, for spectroscopy of the cosmic diffuse supernova
neutrino background, as a detector for the next galactic
supernova, and to search for proton
decay~\citep{Undagoitia:12uu}.
Present design studies for LENA assume
50~kt of liquid PXE scintillator that would provide a geoneutrino rate of
roughly one thousand events per year, if located on the continental crust, from the dominant
\begin{equation}\label{eq:detect}
\bar\nu_e+p\to n+e^+
\end{equation}
inverse beta-decay reaction.

While liquid scintillator detectors do not provide direct angular
information, indirectly one can retrieve directional information
because the final-state neutron is displaced in the forward
direction. The offset between the $e^+$ and the neutron-capture
locations can be reconstructed, although with large uncertainties.
Therefore, it is natural to study the
requirements for a future large-volume liquid scintillator detector to
discriminate between different geophysical models of the Earth that
differ both by their total neutrino fluxes and their neutrino angular
distributions.

Motivated by the current design
studies for LENA we will consider a 50~kt detector using a PXE-based
scintillator. However, it is difficult to locate the
neutron-capture event on protons because a single 2.2~MeV gamma is
released that travels on average 22.4~cm before its first Compton
interaction. Therefore, the event reconstruction is relatively
poor. 

For the geoneutrino flux we will consider a continental and an oceanic location. In each case we will use a reference model and exotic cases
with an additional strong neutrino source in the Earth's core.

We begin in Sec.~\ref{sec:LENA} with a discussion of the
principle of geoneutrino detection in large-volume scintillator
detectors as well as possible scintillator properties.  In
Sec.~\ref{sec:geomodels} we introduce our geophysical models.  In
Sec.~\ref{sec:montecarlo} we turn to the main part of our work, a
Monte-Carlo study of the discriminating power of the LENA detector
between different geophysical models and conclude in
Sec.~\ref{sec:conclusions}.

\section{Geoneutrino detection}
\label{sec:LENA}

\subsection{Directional information from neutron displacement}

In a scintillator detector, geoneutrinos are measured by the inverse
beta-decay reaction Eq.~(\ref{eq:detect}) with an energy threshold of
1.8~MeV. The cross section is
\begin{equation}
\sigma=9.52 \times 10^{-44}~\,{\rm cm}^2\,\,
\frac{E_+}{\rm MeV} \frac{p_+}{\rm MeV}
\end{equation}
where $E_+$ is the total energy of the positron and $p_+$ its
momentum. In this equation the recoil energy ~\citep{vogel} has been neglected, which introduces an error of $\sim 1\%$. The visible energy $E_{\rm vis}=E_++m_e$ always exceeds
1~MeV because the positron annihilates with an electron of the
target. By measuring the visible energy one can determine the neutrino
energy as $E_\nu \approx E_{\rm vis} + 0.8~{\rm MeV}$ because the
kinetic energy of the neutron is typically around 10~keV and thus
negligible.  After thermalization the neutron is captured by a
nucleus, thus tagging the inverse beta decay reaction.

Kinematics implies that the neutron is scattered roughly in the
forward direction with respect to the incoming neutrino~\citep{vogel},
this being the key ingredient for obtaining directional information.
 The average displacement between
the neutron and positron events is then theoretically found to be
about 1.7~cm~\citep{vogel}.

The reactor experiment CHOOZ, using a Gd-loaded scintillator, has
measured an average neutron displacement from the $e^+$ event of
$1.9 \pm 0.4$~cm~\citep{chooz}. However, once the neutron has been
thermalized by collisions with protons, it diffuses some distance
before being captured so that the actual displacement varies by a
large amount for individual events. In a PXE based scintillator the
average time interval until capture on a proton is 180~$\mu$s,
leading to an uncertainty $\sigma$ of the displacement of about 4~cm
for the x-, y- and z-direction~\citep{vogel}. With Gd loading
$\sigma$ is reduced to approximately 2.4~cm~\citep{vogel} because the
neutron diffusion time is much shorter, on average about
30~$\mu$s~\citep{chooz}.

\subsection{PXE-based scintillator}

One option for the proposed LENA detector is to use a scintillator
based on PXE (phenyl-o-xylylethane, C$_{16}$H$_{18}$). PXE has a high
light yield, it is non hazardous, has a relatively high flashpoint of
145$^\circ$C, and a density of 0.985 g/cm$^3$ \citep{Back:2004zn}. A
possible admixture of dodecane (C$_{12}$H$_{26}$) increases the number
of free protons and improves the optical properties.  A blend of
20\%~PXE and 80\%~dodecane shows a decrease in light yield of about
20\% relative to pure PXE, an attenuation length of about 11~m and an
increase in the number of free protons by~25\%~\citep{wurm}.

In this paper we consider a detector with a total volume of about
$70 \times 10^3$~m$^3$. This could be realized with a cylindrical
detector of 100~m length and 30~m diameter.  An outer water
Cherenkov detector with a width of 2~m acts as a muon veto. In order
to shield against external gamma and neutron radiation a fiducial
volume of about $42 \times 10^3~{\rm m}^3$ with a total number of
$2.5 \times 10^{33}$ free protons as target can be realized using a
scintillator mixture as mentioned above with 20\%~PXE and
80\%~dodecane. In Monte-Carlo calculations the light yield of events
in LENA has been estimated ~\citep{Undagoitia:12uu}. For events in
the central detector region the yield $N_{\rm pe}$, measured in
photo-electrons (pe) per MeV energy deposition, can be expressed as
$N_{\rm pe} \approx 400~{\rm pe/MeV} \times c$, where $c$ is the
optical coverage which depends on the number and aperture of the
photomultiplier tubes (PMTs).  A maximal coverage $c_{\rm max}
\approx 0.75$ can not be exceeded so that we assume the maximal
light yield to be around 300~pe/MeV.  For instance, the use of
12,000 PMTs with a diameter in aperture of 50~cm would result in an
optical coverage of about 30\% and a light yield $N_{\rm pe} \simeq
120$~pe/MeV.  This can be obtained either by using PMTs like in the
Super-Kamiokande experiment or by smaller PMTs equipped with light
concentrators as were developed for the Counting Test Facility
(CTF) at the Gran Sasso underground laboratory~\citep{conc}.  For
events off the axis of the cylinder the light yield would be
enhanced.  Hence, low-energy spectroscopy even in the sub-MeV region
should be possible in LENA.

For a detection of the positron-neutron displacement the ability of
the detector to locate the absorption position of both particles is
crucial. The experimental reconstruction of both events is possible
by analyzing the arrival times and the number of photons in each
individual PMT. The position uncertainty depends on the total yield
of registered photo-electrons. In the CTF, the measured position
uncertainty was around 10~cm in each direction for events with 300
photo-electrons and it was shown that the uncertainty scales with
the inverse square-root of that number~\citep{ctf}, as the emission time dispersion of the scintillator is considerably shorter than the photon transient time through the detector. For the 
following discussion we assume, that the scattering length of the scintillator 
is considerably larger than the radius of the detector cyclinder.
  Therefore, we
will use a Gaussian distribution for the uncertainty of the positron
event reconstruction with equal width in each direction~of
\begin{equation}\label{eq:sigma_positron}
\sigma_{e^+}=10~{\rm cm}~\left(\frac{300~{\rm pe/MeV}}{N_{\rm pe}}\,\,
\frac{1~{\rm MeV}}{E_{\rm vis}}\right)^{1/2}
\end{equation}
where $N_{\rm pe}$ is the light yield and $E_{\rm vis}$ the visible
energy released by the positron.

\begin{figure}
\begin{center}
\includegraphics[width=0.6\textwidth]{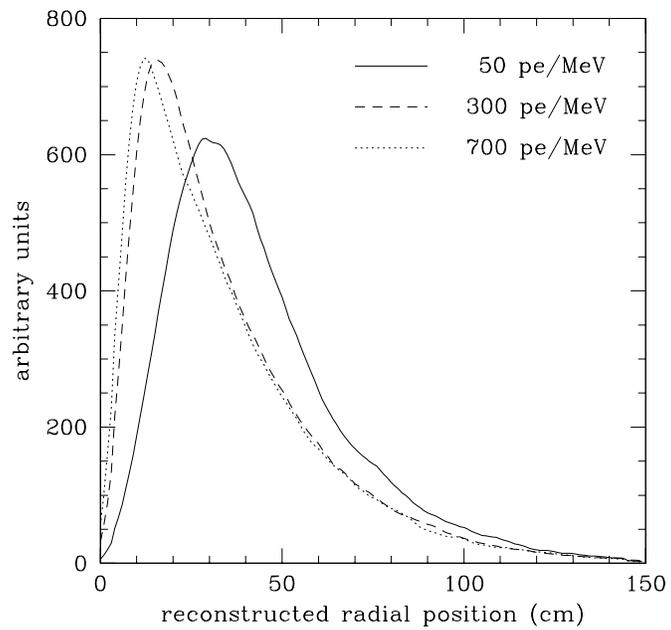}
\caption{\label{fig:1} Monte-Carlo simulation of the radial
distribution of a 2.2~MeV $\gamma$-quantum in an unloaded PXE
scintillator. The curves are for light yields of $N_{\rm pe}=50$,
300 and 700~pe/MeV as indicated.}
\end{center}
\end{figure}

In PXE-based scintillators the neutron is captured by a proton with
nearly 100\% efficiency within an average time interval of about
180~$\mu$s, subsequently emitting a 2.2~MeV gamma.  This photon has a
mean free path of 22.4~cm before its first Compton scattering so that
the event reconstruction is much more uncertain than for the positron
event. We have simulated this case by taking into account multiple
Compton scatterings of the 2.2~MeV gamma.  The position of each gamma
emission, representing the position of the neutron capture, is
reconstructed by composing the energy-weighted sum of each Compton
scattering event, taking into account the instrumental resolution.
The distribution of the reconstructed position in each direction
follows roughly a Lorentzian form. In Fig.~\ref{fig:1} we show the
radial distribution of the reconstructed positions of these events for
light yields of $N_{\rm pe}=50$, 300 and 700~pe/MeV. Increasing the
light yield does not significantly narrow the distribution because its
width is dominated by the large Compton mean free path of the 2.2~MeV
photon.  With reduced light yield the position of the maximum as well
as the mean value of the distribution shifts towards larger
values. This is caused by the increased uncertainty of the
instrumental resolution.

\subsection{Backgrounds}

KamLAND has reported 152 events in the energy region relevant for
geoneutrinos within a measuring time of 749 days and $3.5 \times
10^{31}$ target protons.  From these events $127 \pm 13$ are due to
background~\citep{geokam}. The most relevant background for the
KamLAND site is reactor antineutrinos ($80.4 \pm 7.2$ events). For
the LENA detector positioned in the underground laboratory CUPP (Centre for Underground Physics in Pyh\"asalmi) in
 Finland (longitude: 26$^\circ$ 2.709' E,latitude: 63$^\circ$ 39.579' N,
1450 m of rock (4060 m.w.e.)) this background would be reduced by a factor $\simeq$~12, as
the site is far away from reactors.  Hence we expect for LENA at CUPP
a reactor background rate of about 687 events per year in the relevant
energy window from 1.8~MeV to 3.5~MeV. Assuming a reactor run time of 100\% this rate would increase by 15\% to 790 events. This background can be
subtracted statistically using the information on the entire reactor
neutrino spectrum up to $\simeq$~8 MeV.

Another important background for KamLAND is induced by radio
impurities. A large concentration of the long-lived isotope
$^{210}$Pb is present in the KamLAND scintillator. In the decay
chain of $^{210}$Pb the $\alpha$-emitting isotope $^{210}$Po is present.
Thus the reaction $^{13}$C$(\alpha,n)^{16}$O can occur, mimicking
the signature of geoneutrinos due to neutron scattering on protons
and the subsequent neutron capture. The number of these background
events in KamLAND is estimated to be $42\pm 11$~\citep{geokam}.
However, with an enhanced radiopurity of the scintillator, the
background can be significantly reduced. Taking the radio purity
levels of the CTF detector, where a $^{210}$Po activity of
$35\pm12/\rm{m^3d}$ in PXE has been observed~\citep{Back:2004zn},
this background would be reduced by a factor of about 150 compared
to KamLAND and would account for less than 10 events per year in the
LENA detector.

An additional background that imitates the geoneutrino signal is due
to $^9$Li, which is produced by cosmic muons in spallation reactions
with $^{12}$C and decays in a $\beta$-neutron cascade.  Only a small
part of the $^9$Li decays falls into the energy window which is
relevant for geoneutrinos. KamLAND estimates this background to be
$0.30 \pm 0.05$~\citep{geokam}.  At CUPP the muon reaction rate would
be reduced by a factor $\simeq 10$ due to better shielding and this
background rate should be at the negligible level of $\simeq$~1 event
per year in LENA.

\section{Models of the Earth}
\label{sec:geomodels}
In order to obtain realistic Earth models we use the Bulk Silicate Earth model (McDonough and Sun 1995) abundances for radioactive elements, particularly the reference values derived by \citep{mcfl}, and follow the discussion in~\citep{geonu} to generate angle dependent flux spectra.  For an experiment located on a
continent we have assumed a thickness of 50~km for the crust, implying
a total neutrino flux in our Reference Model of
$4.2\times10^6$~cm$^{-2}$~sec$^{-1}$ from uranium and
$4.1\times10^6$~cm$^{-2}$~sec$^{-1}$ from thorium decays.  For an
oceanic site we have chosen the crust to be rather thick (50~km), but
not included any sediments. If one wanted to determine the mantle
contribution, the oceanic crust would be a background to the
measurement so that the assumption of a thick oceanic crust is
conservative. The neutrino fluxes in this case are
$1.25\times10^6$~cm$^{-2}$~sec$^{-1}$ from uranium and
$0.88\times10^6$~cm $^{-2}$~sec$^{-1}$ from thorium decays.

Besides our reference model we get inspiration from the discussions on additional radioactivity in the core~\citep{hern1,hern3,french} and consider two highly speculative models:

\begin{itemize}
\item[(A)] Fully radiogenic model with additional uranium and
thorium in the core, accounting for 20~TW additional heat production.
(Integrated neutrino flux increase of about 32\% relative to
the reference model in a continental location,
and 116\% in an oceanic location.)
\item[(B)] Same as (A) except with 10~TW in the core. (Flux increase of
16\% and 58\%, respectively.)

\end{itemize}

\begin{table}
\caption{Annual event rates for $2.5\times10^{33}$ target protons.
Flavor oscillations have been included with a global reduction
factor of~0.57.} \label{tab:fermi}
\medskip

\begin{tabular}{lll}
Model& Continental Crust& Oceanic Crust\\
\hline
Reference Model&        $1.02\times 10^3$ &  0.29$\times 10^3$\\
(A) 20 TW core&         1.35$\times 10^3$ & 0.62$\times 10^3$\\
(B) 10 TW core&         1.19$\times 10^3$ &  0.45$\times 10^3$\\
\end{tabular}

\end{table}

To obtain the event rate neutrino flavor
oscillations have to be accounted for by including a global
$\bar\nu_e$ survival-probability factor of~0.57 as measured by
KamLAND~\citep{kamland1}. Matter effects for oscillations are not
important because of the small geoneutrino energies. Moreover, for
geoneutrino energies of 1.8--3.2~MeV and $\Delta
m^2=7.9\times10^{-5}~{\rm eV}^2$ the vacuum oscillation length is
57--101~km. Including distance and energy dependent survival
probabilities is a negligible correction to a global reduction
factor~\citep{fermi}. The annual event rates corresponding to our
models, including the reduction factor, are shown in
Table \ref{tab:fermi} for a 50 kton detector with a fiducial volume
corresponding to $2.5\times 10^{33}$ protons.

Up to now we have assumed that the exotic heat source in the Earth's
core is caused by uranium and thorium decays, i.e.~the neutrino
spectrum from this additional source was taken to be identical with
the geoneutrino spectrum from the crust and mantle. However, the
possibility of a natural reactor in the Earth's core
(``georeactor'') has been discussed in the
literature~\citep{hern1,hern3}. In this case the neutrino flux could
be similar to that from an ordinary power reactor with energies
reaching up to about 8~MeV.  With this assumption the total georeactor neutrino flux can be estimated to be $\Phi_\nu
\simeq 1.9 \times 10^{23}~{\rm s}^{-1}$ for a thermal power of 1~TW.
Taking into account neutrino oscillations, the distance to the
center of the Earth, and the detection cross section we calculate an
event rate of about $210~{\rm y}^{-1}~{\rm TW}^{-1}$ in LENA. At
Pyh\"asalmi one would observe about 2,200 events per year due to
neutrinos from nuclear power plants.  Assuming a systematic uncertainty for the neutrino flux from the power plants of 6.5\%, as suggested in \cite{kamland1}, we conclude that LENA will be able to identify a georeactor of $\ge 2 \,{\rm TW}$ after one year of measurement with a $3\sigma$ significance.

\section{Monte-Carlo Study}
\label{sec:montecarlo}

To study the power of directional discrimination of a large
liquid-scintillator detector we have performed a Monte-Carlo
simulation of a large number of geoneutrino events and the
corresponding directional reconstruction.  We have assumed that the
detector response is independent of the event location, i.e.\ only the
spatial separation between the event $\bar\nu_e+p\to n+e^+$ and the
location of neutron capture is relevant. However, as pointed out in
 Sec.~\ref{sec:LENA}, we consider a position resolution of point-like events
located at the central axis of the detector. We have assumed that, on average, the
neutron capture point is displaced by 1.9~cm in the forward direction
relative to the $e^+$ event in agreement with the CHOOZ
measurement~\citep{chooz19}. Moreover, we have assumed that neutron
diffusion before capture causes a Gaussian distribution around this
mean value with a width $\sigma_x=\sigma_y=\sigma_z=4.0$~cm for an
unloaded PXE-based scintillator as described in Sec.~\ref{sec:LENA}.

In addition to this distribution, the main uncertainty originates from
the reconstruction of both events. For the positron event we have
assumed that the reconstructed location follows a Gaussian
distribution with a width given by Eq.~(\ref{eq:sigma_positron}).  The
actual spread of relevant visible energies is small so that we have
always used $E_{\rm vis}=1.4$~MeV as a typical value.  For an unloaded
scintillator, the reconstruction of the neutron event introduces an
even larger uncertainty; we have used a distribution as in
Fig.~\ref{fig:1} appropriate for the given light yield. 

We conclude that, given the
relatively poor angular reconstruction capability of scintillator
detectors, the only angular-distribution information that can be
extracted is the slope of the geoneutrino distributions. Put another way, one can extract the total event
rate and the dipole contribution of the angular distribution, whereas
a determination of higher multipoles is unrealistic.  Therefore, we
write the reconstructed zenith-angle distribution in the form
\begin{equation}
\frac{d\dot N}{d\cos\theta}=\dot
N\,\left(\frac{1}{2}+p\,\cos\theta\right)
\end{equation}
where the event rate $\dot N$ and the coefficient $p$ are the two
numbers that characterize a given configuration of geophysical model
and detector type.

The event rates for $2.5 \times
10^{33}$ target protons and different geophysical models have
already been reported in Tab.~\ref{tab:fermi}. What remains to be
determined by means of a Monte-Carlo simulation are the
corresponding coefficients $p$ and their uncertainty.  In
Tab.~\ref{tab:dipole} we show the results for $p$ for different
cases, always assuming a light yield of 300~pe/MeV. The uncertainty
$\sigma_p$ of the measured $p$ value scales with the inverse square
root of the number of events $N$ so that $s_p=\sigma_p\sqrt{N}$ is a
quantity independent of $N$. The value of $s_p$ can be derived
analytically for $p=0$, yielding
\begin{equation} \label{eq:s}
s_p=\frac{\sqrt{3}}{2}=0.866,
\end{equation}
which is valid for all $p\ll1$. We have checked with our Monte Carlo
that Eq.~(\ref{eq:s}) indeed applies to all $p$ values of interest
to us.

\begin{table}
\caption{Coefficient $p$ for the reconstructed zenith-angle
  distribution for different Earth models and different detector
  types, always assuming a light yield of
  300~pe/MeV.}\label{tab:dipole}
\medskip

\begin{tabular}{lll}
Model&\multicolumn{1}{l}{Coefficient $p$ for scintillator detectors}\\
\hline
{\bf Continenal Crust}\\
\quad Reference Model& 0.0283   \\
\quad (A) 20 TW core& 0.0377    \\
\quad (B) 10 TW core& 0.0333    \\
{\bf Oceanic Crust}\\
\quad Reference Model&0.0468    \\
\quad (A) 20 TW core& 0.0597   \\
\quad (B) 10 TW core& 0.0560    \\
\end{tabular}

\end{table}

The number of events it takes to distinguish at the $1\sigma$ level
between an isotropic event distribution ($p=0$) and the actual
coefficient is given by $N_{1\sigma}=(s_p/p)^2=(3/4)\,p^{-2}$. For our
reference model at a continental site we find $N_{1\sigma}\approx 500$
events, for an oceanic site about 200 events. In order to distinguish
a geophysical model $i$ from model $j$ at the $1\sigma$ level, the
required number of events is
\begin{equation}
N_{1\sigma}=\frac{2s_{p}^2}{(p_i-p_j)^2}=\frac{3}{2}\,
\frac{1}{(p_i-p_j)^2}.
\end{equation}
A detection at the $n\sigma$ level requires $n^2$ times more events.

\begin{figure}
\begin{center}
\includegraphics[width=0.6\textwidth]{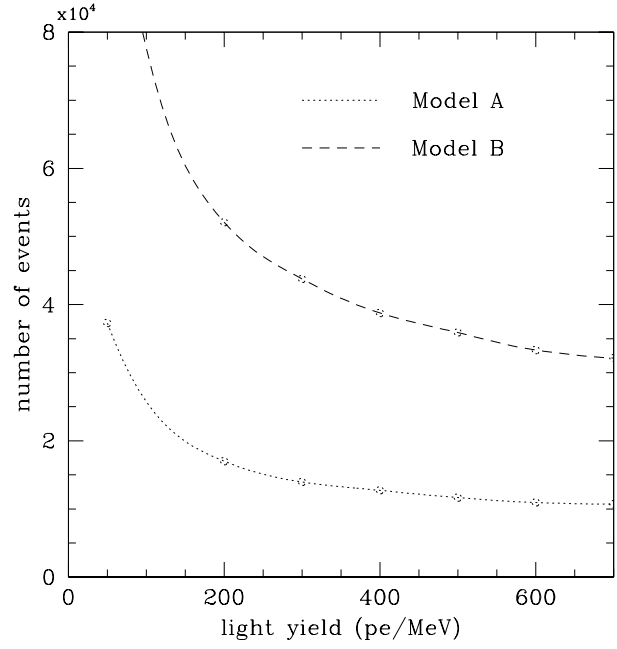}
\includegraphics[width=0.6\textwidth]{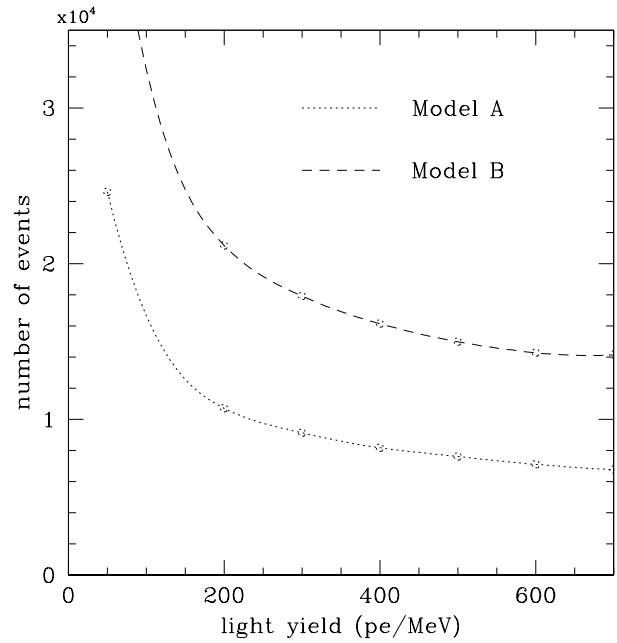}
\end{center}
\caption{Number of events needed to distinguish between models A, B and the continental-crust reference model at $1\sigma$
significance. The points correspond to the values calculated with the
Monte Carlo. {\em Upper panel:} Continental Crust. {\em Lower
panel:} Oceanic Crust} \label{fig:6/7}
\end{figure}

In the same way as for Tab.~\ref{tab:dipole} we have calculated the
slope $p$ for different light yields of the scintillator and have
determined the number of events it takes to distinguish each of the
exotic models from the reference case. In Fig.~\ref{fig:6/7} we
display $N_{1\sigma}$ for these cases and for both the continental- and oceanic-crust
situation as a function of the light yield $N_{\rm pe}$.  

Of course, the time required to achieve this discriminating power
depends on the detector size. For our fiducial volume with $2.5
\times 10^{33}$ target protons as in LENA one needs to scale with
the event rates shown in Tab.~\ref{tab:fermi}. In a
continental-crust location, all models produce an event rate of
roughly 1000 events per year, in full agreement with the KamLAND
measurement~\citep{geokam}. Therefore, even with optimistic
assumptions a 50~kt detector would need several decades for
distinguishing in a meaningful way between different geophysical
models on the basis of the angular event distribution. Moreover,
detector backgrounds should be included in a realistic assessment.

\section{Conclusions}
\label{sec:conclusions}

A future large-volume scintillator detector such as the proposed 50~kt
LENA would provide a high-statistics measurement of the geoneutrino
flux. The event rate would depend strongly on the detector location,
notably on whether an oceanic site such as Hawaii is chosen where a
reference event rate of about 300 per year (50~kt scintillator) is
expected or a continental site such as the Pyh\"asalmi mine in Finland
where the reference rate would be about 1000 per year. Therefore, the
total geoneutrino flux could be measured with high significance and would
allow one to distinguish between different Earth models.

The forward displacement of the neutron in the inverse beta decay
detection reaction provides directional information on the geoneutrino
flux. We have studied if this effect can be used to distinguish
between different geophysical models, notably if one could diagnose a
strong exotic energy source in the Earth's core under the assumption
that its neutrino spectrum is identical with that emitted by the crust
and mantle. While a deviation from an isotropic flux can be
ascertained with high significance, we find that a 50~kt detector is
too small to distinguish between different geophysical models on the
basis of the directional information alone, except perhaps for extreme
cases and optimistic assumptions about the detector performance.

In our study we have only used the neutrino flux from the Earth,
ignoring the contribution from power reactors because it depends
strongly on location. For example, in Pyh\"asalmi the neutrino flux
from power reactors adds roughly 60\% to the counting rate in the
energy window relevant for geoneutrinos. This contribution is not
negligible, but it does not change our overall conclusions.

We have also estimated the sensitivity of a LENA- type detector for
determining a hypothetical georeactor in the Earth's core. As a
possible location the CUPP underground laboratory in Pyh\"asalmi
(Finland) was chosen and the background due to nuclear power plants
was calculated. At CUPP a 2~TW georeactor could be identified at a
statistical level of 3$\sigma$ after only one year of
measurement.

In summary, large-volume scintillator detectors of the next generation
will be extremely useful to study the interior of the Earth in the
``light of neutrinos.'' However, the prime information will be the
total geoneutrino flux and its spectrum. It would be extremely
challenging to use the directional information alone to distinguish
between different geophysical models.

\begin{acknowledgments}
  We thank E.~Lisi for crucial discussions of an earlier version of
  this paper.  Partial support by the Maier-Leibnitz-Laboratorium
  (Garching), the Virtual Institute for Dark Matter and Neutrinos
  (VIDMAN, HGF), the Deutsche Forschungsgemeinschaft under Grant
  No.~SFB-375 and the European Union under the ILIAS project, contract
  No.~RII3-CT-2004-506222, is acknowledged.
\end{acknowledgments}



\begin{thebibliography}{99}
\bibitem[Alimonti {\it et al.}~1998]{ctf}
Alimonti G.~{\it et al.}(1998) [Borexino Collaboration],
  ``A large-scale low-background liquid scintillation detector:
  The Counting Test Facility at Gran Sasso'',
  Nucl. Instrum. Meth. A {\bf 406} (1998) 411.

\bibitem[Apollonio {\it et al.}~1999]{chooz}
  Apollonio M.~{\it et al.}  [CHOOZ Collaboration] (1999),
``Determination of neutrino incoming direction in the CHOOZ experiment  and
  its application to supernova explosion location by scintillator detectors,''
  Phys.\ Rev.\ D {\bf 61}, 012001 (2000)
  [arXiv:hep-ex/9906011].


\bibitem[Apollonio {\it et al.}~2003]{chooz19}
  Apollonio M.~{\it et al.} (2003),
  ``Search for neutrino oscillations on a long base-line at the CHOOZ  nuclear
  power station,''
  Eur.\ Phys.\ J.\ C {\bf 27}, 331 
  [arXiv:hep-ex/0301017].


\bibitem[Araki {\it et al.}~2004]{kamland1}
  Araki T.~{\it et al.}  [KamLAND Collaboration] (2004),
 ``Measurement of neutrino oscillation with KamLAND: Evidence of spectral
  istortion,''
 Phys.\ Rev.\ Lett.\  {\bf 94}, 081801 
  [arXiv:hep-ex/0406035].



\bibitem[Araki {\it et al.}~2005]{geokam}
  Araki T.~{\it et al.} (2005),
  ``Experimental investigation of geologically produced antineutrinos with
  KamLAND,''
  Nature {\bf 436} 499.


\bibitem[Back {\it et al.}~2004]{Back:2004zn}
 Back H.~O.~{\it et al.}  [Borexino Collaboration] (2004),
  ``Phenylxylylethane (PXE): A high-density, high-flashpoint organic liquid
  scintillator for applications in low-energy particle and astrophysics
  experiments,''
  arXiv:physics/0408032.



\bibitem[Fields and Hochmuth 2004]{geonu}
Fields B.~D.~and Hochmuth K.~A.~(2004),
  ``Imaging the earth's interior: The angular distribution of terrestrial
  neutrinos,''
  arXiv:hep-ph/0406001, accepted for publication in Earth, Moon and Planets

\bibitem[Fiorentini {\it et al.}~2003]{fermi}
 Fiorentini G.~, Lasserre T.~, Lissia M.~, Ricci B.~ and Schonert S.~(2003),
  ``KamLAND, terrestrial heat sources and neutrino oscillations,''
  Phys.\ Lett.\ B {\bf 558}, 15 
  [arXiv:hep-ph/0301042].


\bibitem[Herndon 1993]{hern1}
 Herndon J~.M.~(1993), J. Geomagn. Geoelectr. 45, 423 

\bibitem[Herndon 2003]{hern3}
 Herndon J.~M.~(2003), 
``Nuclear Georeactor Origin of Oceanic Basalt He–3/He–4, Evidence, and Implications,''
PNAS, March 18 ,
Vol. 100, No. 6, pp.~3047-3050

\bibitem[Labrosse {\it et al.}~2001]{french}
Labrosse S.~{\it et al.}~(2001), Earth Planet. Sci. Lett. 190, 111 

\bibitem[Mantovani {\it et al.}~2003]{mcfl}
Mantovani F.~, Carmignani L.~, Fiorentini G.~ and Lissia M.~(2003),
 ``Antineutrinos from the earth: The reference model and its  uncertainties,''
  Phys.\ Rev.\ D {\bf 69}, 013001
  [arXiv:hep-ph/0309013].


\bibitem[McDonough and Sun 1995]{macsun}
 McDonough W.~F.~ and Sun S.~-s.~(1995), "The composition of the Earth'',Chem. Geol. 120, 223



\bibitem[Oberauer {\it et al.}~2005]{Oberauer:2005kw}
Oberauer L.~, von Feilitzsch F.~ and Potzel W.~ (2005),
  ``A large liquid scintillator detector for low-energy
  neutrino astronomy,''
  Nucl.\ Phys.\ Proc.\ Suppl.\  {\bf 138} 108.

\bibitem[Oberauer {\it et al.}~2004]{conc}
Oberauer  L.~, Grieb C.~,von Feilitzsch  F.~,Manno  I.~(2004),
  ``Light concentrators for Borexino and CTF,''
  Nucl.\ Instrum.\ Meth.\ A {\bf 530} 453
  [arXiv:physics/0310076].

\bibitem[Vogel and Beacom 1999]{vogel}
  Vogel P.~and Beacom J.~F.~(1999),
  ``The angular distribution of the neutron inverse beta decay, anti-nu/e + p
  $\to$ e+ + n,''
  Phys.\ Rev.\ D {\bf 60} 053003
  [arXiv:hep-ph/9903554].

\bibitem[Marrod\'an Undagoitia {\it et al.}~2005]{Undagoitia:12uu}
Marrod\'an Undagoitia   T.~{\it et al.} (2005),
  ``Search For The Proton Decay P $\to$ K+ Anti-Nu In The Large Liquid
  Scintillator Low Energy Neutrino Astronomy Detector Lena,''
  Phys.\ Rev.\ D {\bf 72}, 075014 (2005)
  [arXiv:hep-ph/0511230].


\bibitem[Wurm 2005]{wurm}
  Wurm M.~(2005),
  ``Untersuchungen zu den optischen Eigenschaften eines
  Fl\"ussigszintillators und zum Nachweis von Supernovae Relic
  Neutrinos mit LENA'', Diploma thesis, TU M\"unchen, Germany (2005).





\end{thebibliography}
\end{document}